\documentclass[twocolumn,superscriptaddress,amsmath,amssymb,showpacs,showkeys]{revtex4}

\usepackage{graphicx}
\usepackage{graphicx,color}
\usepackage{dcolumn}
\usepackage{bm}
\usepackage[colorlinks=true]{hyperref}
\newcommand{\beq}{\begin{equation}} \newcommand{\eeq}{\end{equation}}
\newcommand{\bqa}{\begin{eqnarray}} \newcommand{\eqa}{\end{eqnarray}}

\definecolor{gold}{rgb}{0.75,0.56,0.00}
\definecolor{green}{rgb}{0.00,0.50,0.00}

\newcommand{\ms}[1]{\mbox{\scriptsize #1}}

\begin{document}

\preprint{APS/123-QED}

\title{Coupling rotational and translational motion via a continuous measurement in an optomechanical sphere}

\author{Jason F. Ralph}
 \email{jfralph@liverpool.ac.uk}
 \affiliation{Department of Electrical Engineering and Electronics, University of Liverpool,  Brownlow Hill, Liverpool, L69 3GJ, UK.}
\author{Kurt Jacobs}
 \email{kurt.jacobs@umb.edu}
  \affiliation{U.S. Army Research Laboratory, Computational and Information Sciences Directorate, Adelphi, Maryland 20783, USA.}
  \affiliation{Department of Physics, University of Massachusetts at Boston, Boston, MA 02125, USA} 
\affiliation{Hearne Institute for Theoretical Physics, Louisiana State University, Baton Rouge, LA 70803, USA} 
\author{Jonathon Coleman}
 \email{coleman@liverpool.ac.uk}
 \affiliation{Department of Physics, University of Liverpool,  Oxford Street, Liverpool, L69 7ZE, UK.}

\date{\today}

\begin{abstract}
We consider a measurement of the position of a spot painted on the surface of a trapped nano-optomechanical sphere. The measurement extracts information about the position of the spot and in doing so measures a combination of the orientation and position of the sphere. The quantum back-action of the measurement entangles and correlates these two degrees of freedom.  Such a measurement is not available for atoms or ions, and provides a mechanism to probe the quantum mechanical properties of trapped optomechanical spheres. In performing simulations of this measurement process we also test a numerical method introduced recently by Rouchon and collaborators for solving stochastic master equations. This method guarantees the positivity of the density matrix when the Lindblad operators for all simultaneous continuous measurements are mutually commuting. We show that it is both simpler and far more efficient than previous methods. 
\end{abstract}



\maketitle

\section{\label{sec:sec1}INTRODUCTION}

The optical trapping of atoms, molecules and even very small objects is now an established experimental tool used to explore fundamental physics at nanoscopic length scales~\cite{Ash1970, Man1998, Gan1999, Vul2000, Vul2001, Mil2005, Mar2007, Wil2007,Wil2008,Gen2008,Rom2010,Cha2010,Bar2010,Sin2010,Rom2011b,Che2012,Yin2013}. The development of this technology has been a crucial step in the exploration of the physics of individual atoms, and plays an important role in testing the boundaries of quantum theory \cite{Bas2003,Bas2013,Ber2015}. The ability to levitate a single particle of matter in an electromagnetic field allows the particle to be isolated from many of the environmental effects that would inhibit the experimental investigation of subtle quantum effects \cite{Ber2015}. One particular area of active study which relies on optical trapping is the levitation and cooling of nano-spheres \cite{Rom2011b}, nano-rods \cite{Sti2016,Kuh2015} and other nanostructures \cite{Hoa2016}. By cooling such objects to very low temperatures, systems should approach the quantum mechanical ground states of their motional degrees of freedom. In this regime, one would expect quantum mechanical effects to appear, and any deviations from standard quantum mechanics to be evident \cite{Bas2003,Bas2013,Ber2015}. Probing the ability of quantum mechanics to describe relatively large systems (relative to single atoms) is an important scientific question; one that could have implications for the emergent fields of quantum technology and quantum computing -- as quantum systems grow in size and complexity, any modifications to the fundamental theory may become limiting factors in the ability to construct large-scale systems. 

The present paper has two principal objectives: the introduction of a novel measurement interaction for small optically trapped objects, and the demonstration of the advantages of a numerical integration method proposed by Rouchon~\textit{et al.\ }for stochastic master equations~\cite{Ami2011,Rou2015}. The measurement relies on the fact that the object is spatially extended, which allows one part of it to be localized without necessarily localizing the position of the whole object (that is, without localizing the center of mass). We demonstrate the power of Rouchon's numerical method by employing it for an example system which contains an order of magnitude more quantum states than the qubit-based examples for which it has been tested previously~\cite{Rou2015}. We find that, for the integration of those stochastic master equations for which it applies, Rouchon's method is significantly more stable and more accurate than the Euler-Milstein stochastic integration method~\cite{Mil1995}. 

In this paper we choose as an example, a nanoscopic dielectric sphere held in a one-dimensional harmonic potential formed by optical tweezers~\cite{Ash1970}. For the purposes of modeling the measurement process, we will approximate the system to two degrees of freedom (one translational and one rotational). We show that the effect of the back-action of the quantum measurement is to induce a coupling between the two degrees of freedom, and that this leads to correlations in the motional states of the sphere. The continuous measurement produces a continuous stream of measurement results with a necessarily random component. The dynamics of the sphere induced by the measurement are described by a stochastic master equation (SME), an equation of motion for the density matrix driven by the noise on the measurement results. At a time $t$, the SME provides the observer's complete state-of-knowledge of the system, in the form of the density matrix, $\rho(t)$, based on all of the information provided by the stream of measurement results obtained up to that time~\cite{Jacobs2014}. The evolution of the density matrix under a continuous measurement is often called a {\em quantum trajectory}~\cite{Bel1980,Wis2010}. The effect of the coupling is maximized when the energy scales associated with the two degrees of freedom are similar, and the coupling can be controlled by modulating the measurement interaction, using the information provided by the density matrix to modify the measurement strength. In addition to the environmental interaction that mediates the measurement the sphere may be coupled to other environments which may cause thermalization, dephasing, and other forms of decoherence. 

\begin{figure}[t] 
   \centering
   \includegraphics[width=0.8\hsize]{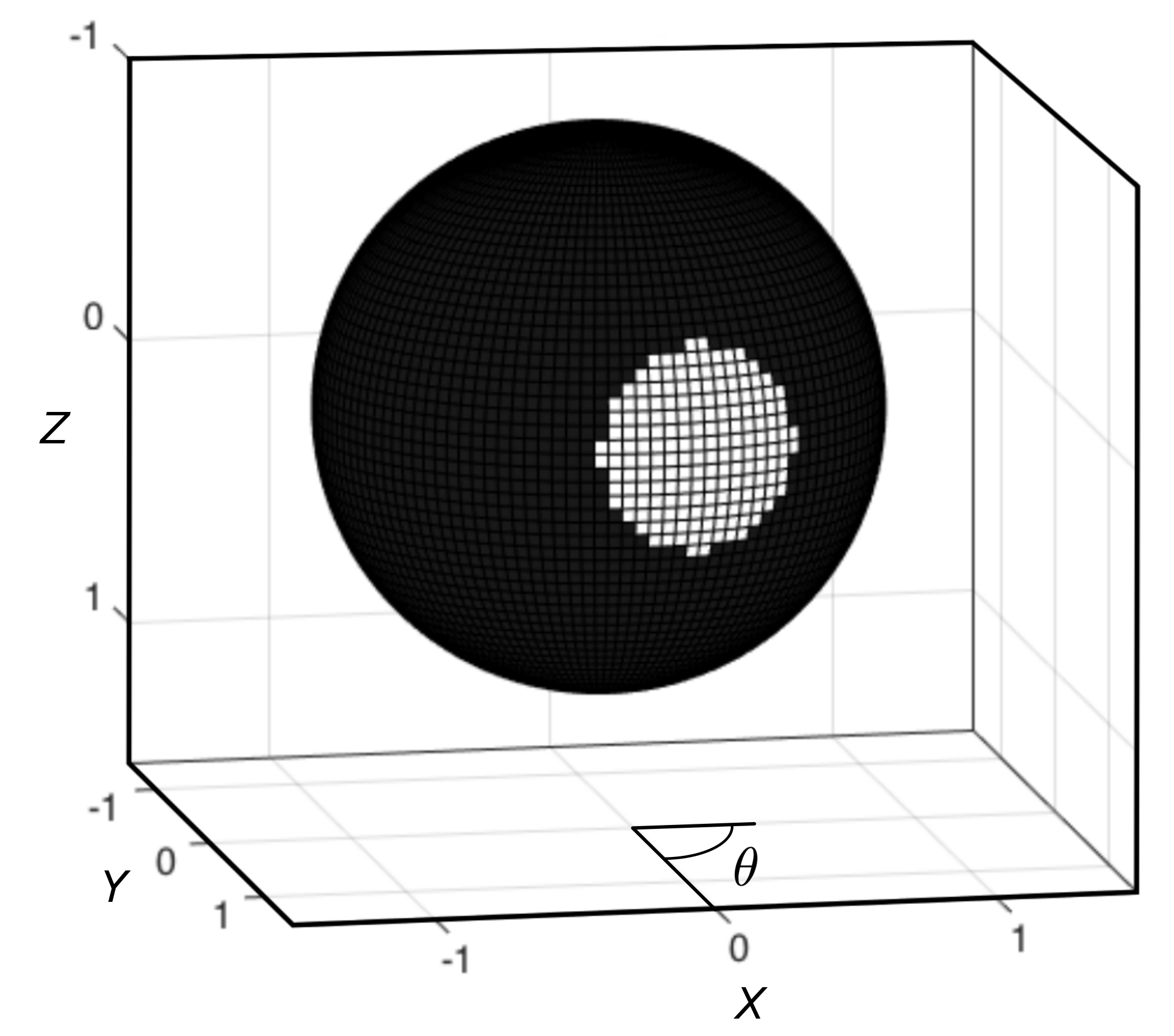} 
   \caption{Schematic representation of the dielectric sphere with fluorescent `spot'. The position of the spot along the $x$-axis is a combination of the position of the center of the sphere along the $x$-axis, $X$, and the angle of the rotation of the sphere in the $xy$-plane, $\theta$. The spot is shown larger than it would be in an experiment.}
   \label{fig:sphere}
\end{figure}

We consider a measurement of the position of a small `spot' of fluorescent material placed on the surface of the dielectric sphere. This measurement could be implemented, for example, by exciting the fluorescence of the spot and imaging the emitted light. A derivation of the usual SME that describes a continuous measurement for this implementation can be found in~\cite{Jacobs2006}. What is novel here is that the position of the spot has two contributions, one from the position of the center of mass of the sphere, and one from its orientation. Thus, we measure a sum of these two degrees of freedom, and this provides the mechanism for entangling them. As an example, in the extreme case of a very strong measurement, the location of the spot would be fixed by the Zeno effect~\cite{Mis1977, Kha1958, Win1961}, allowing the sphere to be in exactly one location for each orientation, and so the two degrees of freedom would be perfectly correlated. A much weaker measurement will generate correlations via a less constrained version of the same mechanism. While we consider here a measurement of a single spot on an otherwise `dark' sphere (akin to the `8 ball' in the billiard game Pool, see Fig.~\ref{fig:sphere}), we note that this approach to joint measurements of position and orientation could be generalized: a selection of different patterns (e.g. `spots' and `stripes') on the sphere might be arranged to produce various types of coupling between translational and rotational states. We also note that this type of measurement could not be performed on a trapped atom or an ion, but might be envisaged for very large molecules, where the position of a spatially localized excitation could be measured within a much larger, extended molecular structure.

The second objective of the present work is to apply a numerical stochastic integration method proposed by Rouchon et al. \cite{Ami2011, Rou2015} to a system which is significantly larger, in terms of its quantum state-space, than that studied previously \cite{Rou2015}. The integration method is specific to stochastic master equations and we confirm here that it offers significant advantages: i) it requires far fewer time increments to obtain an accurate solution when compared to more general numerical integration methods; ii) it is simpler in that each time increment requires fewer operator calculations than comparable methods -- both i) and ii) reduce the simulation time; and, iii) it is much more stable in that, unlike previous methods, it does not produce unphysical density matrices in which the purity is greater than unity.

This paper is structured as follows. We start by introducing the basic optomechnical system, a nanoscopic dielectric sphere levitated by optical tweezers, and the approximations required to model the system as two degrees of freedom (one translational and one rotational). In Sec.~3, we introduce the measurement operator and construct the stochastic master equation for a continuous measurement. We then describe the numerical integration method in Sec.~4 and the numerical results in Sec.~5. Sec.~6 provides a discussion of the control of the coupling through the modulation of the measurement strength, how the properties of the system may provide a route to investigate the boundary between quantum and classical physics, and the effect of the approximations used in realizing the measurements described in the paper. The main conclusions are then summarized in Sec.~7.

\section{\label{sec:sec2}OPTOMECHANICAL SYSTEM}

The optical trap dates back to 1970 with the work of Ashkin \cite{Ash1970}. Optical traps have become common tools in the manipulation of tiny neutral objects, down to individual atoms and molecules. Such systems have a wide range of applications: sensors for measuring tiny forces \cite{Ger2010}, tests of fundamental science and the laws of gravity \cite{Arv2013}, and investigation of the quantum mechanical behavior of macroscopic objects \cite{Bas2003,Bas2013,Ber2015,Rom2011a}. To probe the quantum behavior of nanoscopic objects rather than individual atoms (i.e. objects that, while only tens of nanometers across, contain a relatively large number of atoms), the key is to cool the system down to very low temperatures so that many of the microscopic excitations associated with the internal structure can be neglected \cite{Rom2011b}. 

A standard experimental system is a small dielectric sphere that is held in a harmonic trapping potential generated by optical tweezers, and coupled to an optical cavity field. The cavity field can be used to manipulate its motion (including feedback cooling) and to measure its position \cite{Rom2011b,Yin2013}. The general theory and experimental set-up for such a system are described in references \cite{Rom2011b,Che2012,Yin2013}. The main degrees of freedom of a nanoscopic dielectric sphere are its 3D translational states, its 3D rotational states, and its internal (vibrational) states. The vibrational states can be shown to have significantly higher frequencies than the motional states, and so can be neglected at low temperatures \cite{Rom2011b}. Direct coupling between the translational states, described by the operators $X,Y,Z$, and the rotational states, described by the total angular momentum ${\bf J} = j(j+1)$, $j=0,1,2,...$, and the components $J_x,J_y,J_z$, are very weak and usually neglected~\cite{Rom2011b,Che2012}. The Hamiltonian for this system is given by~\cite{Rom2011b,Che2012} 
\begin{equation}
\hat{H} = \sum_{k=x,y,z}\omega_{t}^{(k)} \hat{a}_k^{\dagger}\hat{a}_k+ \frac{\hat{{\bf J}}^2}{2 I}+ \sum_{\omega_c}\omega_c \hat{b}^{\dagger}(\omega_c)\hat{b}(\omega_c)+H_{ab} , 
\end{equation}
where the frequencies of the (harmonic) optical trapping potential are denoted by $\omega_{t}^{(k)}$, the operators $\hat{a}_k^{\dagger}$ ($\hat{a}_k$) are the corresponding raising (lowering) operators for translational excitations of the center of mass of the sphere in the trap, $I=(2/5) mR^2$ is the moment of inertia ($R$ is the radius and $m$ the mass of the sphere), $\hat{b}_k^{\dagger}(\omega_c)$ ($\hat{b}_k(\omega_c)$) are the raising (lowering) operators for the optical cavity modes with frequency $\omega_c$, and $H_{ab}$ is the coupling between the center of mass modes and the cavity modes. We have set $\hbar=1$ for simplicity.

The translational motion associated with the sphere's center of mass within the potential generated by the optical tweezers can be represented by a 3D harmonic trap, where -- for convenience -- we assume that the motion in the $Y$ and the $Z$ directions is sufficiently constrained for the problem to be reduced to one spatial dimension (motion in the $X$ direction). In practice, the potential generated by the optical tweezers is asymmetric and one motional state will have a lower frequency than the other two, and often it is this axis which is aligned to the optical cavity \cite{Rom2011b}. The rotational states are more complicated, because in addition to the familiar atomic angular momentum states for a total angular momentum ${\bf J}$, a rigid body has an additional angular momentum $J'$ corresponding to the angular momentum component in body fixed axes (as opposed to the space fixed axes components $J_x,J_y,J_z$) \cite{Edm1996}. This means that each of the angular momentum components corresponding to $J_x,J_y,J_z$ are multiply degenerate, giving a total number of angular momentum states of $(2j+1)^2$ rather than the more familiar $(2j+1)$ states. However, we will reduce the rotational degrees of freedom to one by making an approximation corresponding to ${\bf J}\rightarrow\infty$ \cite{Car1968} so that we are left with one angular momentum operator $J_z$ and one angular variable $\theta$. In this case, the multiply degenerate angular states corresponding to the body fixed axes do not couple to the one remaining angular degree of freedom and can be neglected for simplicity. With these approximations, and dropping the explicit dependence on $X$ and $\omega_c$, the standard optomechanical Hamiltonian for the system is given by \cite{Blo2008},
\begin{equation}
   \hat{H} = \omega_{t}\hat{a}^{\dagger}\hat{a}+ \omega_c \hat{b}^{\dagger}\hat{b}+g_0\hat{a}^{\dagger}\hat{a}(\hat{b}^{\dagger}+\hat{b})+\frac{\hat{J_z}^2}{2 I}
\end{equation}
where the coupling Hamiltonian is expressed in terms of a coupling constant $g_0$, a number operator $\hat{a}^{\dagger}\hat{a}$, and an cavity position operator $(\hat{b}^{\dagger}+\hat{b})$,
$$
H_{ab}=g_0\hat{a}^{\dagger}\hat{a}(\hat{b}^{\dagger}+\hat{b})
$$
For the sake of simplicity, we will also remove the explicit optical cavity from the Hamiltonian. However, we will later couple the center of mass motion to a dissipative thermal bath to represent all of the unprobed (i.e. un-measured) degrees of freedom present in the system. The reduced Hamiltonian for the one dimensional center of mass motion and quasi-one dimensional rotational motion, is then given by,
\begin{equation}
\hat{H} \simeq \omega_t \hat{a}^{\dagger}\hat{a}+ \frac{\hat{J_z}^2}{2 I}
\end{equation}
The parameters that we use for the system are based on those given in reference~\cite{Rom2011b}: a fused silica sphere with a density of $2201$ kg/m$^3$, but with a significantly smaller sphere radius of around $1.5-3$ nm (corresponding to approximately 10-20 atomic radii), and optical tweezers with a resonant frequency of $\omega_t = 2\pi\times 135$ kHz. The size of the sphere is smaller to match the relative energy scales of the translational and rotational energies of the tweezers and the angular momentum of the sphere -- rotational energies, $E_{Rot} \simeq \hbar^2/(2 I)$, and translational energies, $E_{Tr}\simeq \hbar\omega_t$. This maximizes the effect of the coupling/energy-exchange between the two degrees of freedom. An alternative would be to use a larger sphere and vary the frequency of the optical tweezers. 

\section{\label{sec:sec3}MEASUREMENT-INDUCED COUPLING} 

A standard approach to the measurement of a dielectric sphere in an optical trap would be to couple to the sphere's center of mass motion via a laser cavity~\cite{Rom2011b}. The measured position of the sphere can then be used as an input into a feedback cooling mechanism~\cite{Rom2011b,Yin2013}. In this paper, we envisage an alternative measurement, where the sphere has a `bright' spot (possibly due a fluorescent atom or a few fluorescent atoms on its surface) and it is the location of the bright spot that is measured (e.g. via an optical microscope). Using a fluorescent spot would have the advantage that the rate at which the measurement extracts information could be modulated by varying the illumination of the sphere. We will assume that any illumination is at a sufficiently low wavelength and low intensity that recoil of the sphere can be neglected. 

The dynamics of the sphere under the continuous measurement of the spot's position is described by the usual SME for a continuous measurement, namely~\cite{Jacobs2006, Jacobs2014} 
\begin{align}\label{sme_hermitian}
   d\rho_{\ms{c}} = & -i [\hat{H}, \rho_{\ms{c}}] dt - k [\hat{x},[\hat{x},\rho_{\ms{c}}]] dt  \nonumber \\ 
                              &+ \sqrt{2k}(\hat{x} \rho_{\ms{c}} + \rho_{\ms{c}} \hat{x} - 2 \mbox{Tr}[\hat{x}\rho_{\ms{c}}] \rho_{\ms{c}} ) dW , 
\end{align} 
where $\hat{x}$ is the operator of the spot's position, $k$ is the measurement strength, $\rho_{\ms{c}}$ is the density matrix for the sphere, and $dW$ is calculated from the stream of measurement results, $y(t)$, using 
\begin{align}
  dW(t) =  \sqrt{8k} ( dy - \mbox{Tr}[\hat{x}\rho] dt ), 
\end{align} 
with $dy = y(t+dt)-y(t)$. The subscript `c' in our notation for the density matrix denotes the fact that the density matrix is the observer's state-of-knowledge based on the stream of measurement results (the {\em conditioned} density matrix). If the observer discards the measurement results then her state-of-knowledge is instead given by taking the SME and averaging over the stochastic increment $dW$ to obtain the master equation
\begin{align}\label{me_hermitian}
   \dot{\rho} = -i [\hat{H}, \rho]  - k [\hat{x},[\hat{x},\rho]] . 
\end{align} 
Because the spot is on the surface of the sphere, the position of the spot is a function of the position of the center of the sphere, $\hat{X}=\sqrt{\frac{\hbar}{2 m \omega_t}}(\hat{a}^{\dagger}+\hat{a})$, and the angle of rotation of the sphere, $\hat{\theta}$. Specifically we have 
\begin{align}
   \hat{x} = \hat{X}+R \hat{S} , 
   \label{hatx}
\end{align}
where the operator $\hat{S}$ represents the sine of the angular variable $\theta$ and it obeys the commutation relation~\cite{Car1968}  
\begin{align}
    [\hat{S},J_z]=  i \hat{C} . 
\end{align} 
where the $\hat{C}$ operator represents the cosine of the angle. Technically, in order for the measured observable, $\hat{x}$, to be given by Eq.(\ref{hatx}) the sphere would need to be transparent at the wavelength used for the illumination. This is to ensure that the position of the spot could be measured even when on the `back' of the sphere. An alternative would be to have two spots, one on each side, but the operator would then represent $|\sin\theta |$ rather than the sine operator. However, this addition complicates the scenario unnecessarily. 

In addition to the measurement given above we include the effects of the thermal environment on the center of mass motion of the sphere~\cite{Spi1993}. The action of an environment is very similar to that of a continuous measurement; this action is described by adding terms to the master equation that have the Lindblad form, and these are equivalent to a continuous measurement in which the `observable' being measured is a non-Hermitian operator and the results of the measurement are averaged over. In fact, it is in theory possible to turn the environmental interaction into a continuous measurement by monitoring the environment, but this is not always practical.  With the addition of the thermal bath the SME describing the sphere becomes 
\begin{align}
   d\rho_{\ms{c}} = & -i [\hat{H}, \rho_{\ms{c}}] dt - k [\hat{x},[\hat{x},\rho_{\ms{c}}]] dt  \nonumber \\ 
                              &+ \sqrt{2k}(x \rho_{\ms{c}} + \rho_{\ms{c}} x - 2 \mbox{Tr}[\hat{x}\rho_{\ms{c}}] \rho_{\ms{c}} ) dW \nonumber \\ 
                              & + \sum_{r= 2,3} \left\{\hat{L}_{r} \rho \hat{L}^{\dagger}_{r} -\frac{1}{2}\left(\hat{L}^{\dagger}_{r} \hat{L}_{r} \rho + \rho \hat{L}^{\dagger}_{r} L_{r} \right)\right\}dt , 
                              \label{sme2}
\end{align} 
in which 
\begin{align}
\hat{L}_2 = \, \sqrt{\frac{(\bar{n}+1)\omega_t}{Q}} \, \hat{a} ,  \;\;\;\;  \hat{L}_3  = \, \sqrt{\frac{\bar{n}\omega_t}{Q}} \, \hat{a}^{\dagger} . 
\end{align}
Here $Q$ is the quality factor of the harmonic potential and $\bar{n} = [\exp(-\hbar\omega_t/k_{\ms{B}} T) - 1]^{-1}$ where $k_{\ms{B}}$ is Boltzmann's constant and $T$ is the temperature of the environment.  

The thermal environment introduces decoherence, dephasing, and dissipation into the dynamics of the sphere, and in general will cause an initially pure state to evolve into a mixed state. This limits our ability to purify the state of the system via the measurement of $\hat{x}$. At some level all physical systems are coupled to a thermal bath, even if weakly.  We set the quality factor of the harmonic potential to be $Q = 100$ and $T = 5$ $\mu$K to ensure that only a few lowest lying energy states will be populated by the thermal noise. This choice also reduces the computational demands of the simulations by reducing the dimension of the required state-space.

\section{\label{sec:sec4}NUMERICAL METHOD}

All Markovian stochastic master equations for quantum systems can be written in the general form~\cite{Wis2010, Jacobs2014},
\begin{eqnarray}\label{sme1}
d\rho_{\ms{c}}&=&- i \left[\hat{H},\rho_{\ms{c}}\right]dt \nonumber \\
&&+\sum_{r=1}^{m} \left\{ \hat{L}_{r} \rho_{\ms{c}} \hat{L}^{\dagger}_{r} -\frac{1}{2}\left(\hat{L}^{\dagger}_{r} \hat{L}_{r} \rho_{\ms{c}} + \rho_{\ms{c}} \hat{L}^{\dagger}_{r} \hat{L}_{r} \right)\right\}dt   \nonumber \\
&&+ \sum_{r=1}^{m} \sqrt{\eta_r}\left(\hat{L}_{r}\rho_{\ms{c}}+\rho_{\ms{c}} \hat{L}^{\dagger}_{r}-\mathrm{Tr}(\hat{L}_{r}\rho_{\ms{c}}+\rho_{\ms{c}} \hat{L}^{\dagger}_{r}) \right)dW_{r}\nonumber \\
\label{genSME}
\end{eqnarray}
where each operator $\hat{L}_r$ corresponds to a continuous measurement or the action of an unmonitored/unprobed environment. The parameters $\eta_r$ are called the measurement \textit{efficiencies}. We have taken $dW_{r} $ to be real Wiener increments (such that $\langle dW_{r} \rangle =0$ and $dW_{r} dW_{r'}  = \delta_{r,r'}dt$, where $\delta_{r,r'}$ is the Kronecker delta symbol). Setting $\eta_{r} = 1$ describes a perfectly efficient measurement in which all available information is collected and $\eta_r = 0$ correspond to an unmonitored environment. The stream of measurement results for those values of $r$ that are monitored are given by $y_r$ in which 
\begin{align}
   y_r(t+dt) =  y_r(t) + \sqrt{\eta_j}\mathrm{Tr}(\hat{L}_{r}\rho_{\ms{c}}+\rho_{\ms{c}} \hat{L}^{\dagger}_{r}) dt+dW_{r} . 
\end{align}

The SME for our system, Eq.(\ref{sme2}), is obtained from the above general form by setting $m=3$, defining $L_2$ and $L_3$ as above, choosing 
\begin{align}
   L_1 = \sqrt{2k} \hat{x}, 
\end{align}
and setting $\eta_1 = 1$ and $\eta_2=\eta_3 = 0$. 

For a given initial state, a system Hamltonian $H$, and a set of environmental operators $\hat{L}_r$, the SME may be integrated using standard numerical stochastic integration methods, e.g. the Euler-Maruyama method which is weakly convergent to first order or the Euler-Milstein method which is strongly convergent to first order~\cite{Mil1995}. Here we adopt instead a numerical integration method specifically developed for SMEs~\cite{Ami2011,Rou2015} that we will refer to as Rouchon's method.

When all of the operators $\hat{L}_r$ for which $\eta_r \not= 0$ are mutually commuting, Rouchon's method is convergent to first order and it guarantees the positivity of the conditional density matrix (up to numerical rounding errors). It has this property because it is based on the representation of the SME in terms of a positive operator-valued measure (POVM)~\cite{Wis2010,Jacobs2014}. For Rouchon's method the increment to the conditional density matrix for the time step from $t_n = n\Delta t$ to $t_{n+1} = (n+1)\Delta t$ is given by $\Delta \rho_{\ms{c}}^{(n)} = \rho_{\ms{c}}^{(n+1)}- \rho_{\ms{c}}^{(n)}$, where
\begin{equation}\label{sme2}
\rho_{\ms{c}}^{(n+1)}= \frac{\hat{M}_n\rho_{\ms{c}}^{(n)}\hat{M}_n^{\dagger}+\sum_{r=1}^{m}(1-\eta_r)\hat{L}_r\rho_{\ms{c}}^{(n)}\hat{L}_r^{\dagger}\Delta t }
{\mathrm{Tr}\left(\hat{M}_n\rho_{\ms{c}}^{(n)}\hat{M}_n^{\dagger}+ \sum_{r=1}^{m}(1-\eta_r)\hat{L}_r\rho_{\ms{c}}^{(n)}\hat{L}_r^{\dagger}\Delta t \right)}
\end{equation}
and $\hat{M}_n$ is given by 
\begin{widetext}
\begin{eqnarray}\label{Mn1}
\hat{M}_n &=& I-\left(i\hat{H} +\frac{1}{2}\sum_{r=1}^{m} \hat{L}^{\dagger}_{r}\hat{L}_{r}\right)\Delta t  +\sum_{r=1}^{m} \sqrt{\eta_r}\hat{L}_{r}\left(\sqrt{\eta_r}\mathrm{Tr}(\hat{L}_{r}\rho_{\ms{c}}^{(n)}+\rho_{\ms{c}}^{(n)}\hat{L}^{\dagger}_{r})\Delta t +\Delta W_{r} (n)\right)\nonumber\\
&&+\sum_{r,s=1}^{m}  \frac{\sqrt{\eta_r \eta_s}}{2}\hat{L}_{r}\hat{L}_{s}(\Delta W_r(n)\Delta W_s(n)-\delta_{r,s}\Delta t) , 
\end{eqnarray}
\end{widetext}
where the $\Delta W_r$'s are independent Gaussian variables with zero mean and a variance equal to $\Delta t$.  The stochastic increment $\Delta W_r$ is only half order in $\Delta t$, so terms proportional to $(\Delta W_r)^2$ need to be retained~\cite{Mil1995}. 

For comparison the Euler-Milstein increment for a finite time step $\Delta t$ is~\cite{Mil1995,Tal1995},
\begin{widetext}
\begin{eqnarray}\label{sme3}
\Delta \rho_{\ms{c}}&=&- i \left[\hat{H},\rho_{\ms{c}}\right]\Delta t+\sum_{r=1}^{m} \left\{\hat{L}_{r} \rho_{\ms{c}} \hat{L}^{\dagger}_{r} -\frac{1}{2}\left(\hat{L}^{\dagger}_{r} \hat{L}_{r} \rho_{\ms{c}} + \rho_{\ms{c}} \hat{L}^{\dagger}_{r} \hat{L}_{r} \right)\right\}\Delta t+ \sum_{r=1}^{m}  \sqrt{\eta_r}\left(\hat{L}_{r}\rho_{\ms{c}}+\rho_{\ms{c}} \hat{L}^{\dagger}_{r}-\mathrm{Tr}(\hat{L}_{r}\rho_{\ms{c}}+\rho_{\ms{c}} \hat{L}^{\dagger}_{r}) \right)\Delta W_{r}  \nonumber \\
&&+ \sum_{r,s=1}^{m}  \frac{\sqrt{\eta_r \eta_s}}{2}
\left(\begin{array}{l}
\hat{L}_r \hat{L}_s \rho_{\ms{c}}+\rho_{\ms{c}} \hat{L}^{\dagger}_{r} \hat{L}^{\dagger}_{s}+\hat{L}_{s}\rho_{\ms{c}} \hat{L}^{\dagger}_{r}+ \hat{L}_{r}\rho_{\ms{c}} \hat{L}^{\dagger}_{s} \\
-\mathrm{Tr}\left(\begin{array}{l}
\hat{L}_r \hat{L}_s \rho_{\ms{c}}+\rho_{\ms{c}} \hat{L}^{\dagger}_{r} \hat{L}^{\dagger}_{s}+\hat{L}_{s}\rho_{\ms{c}} \hat{L}^{\dagger}_{r}+ \hat{L}_{r}\rho_{\ms{c}} \hat{L}^{\dagger}_{s}
\end{array}\right)\rho_{\ms{c}} \\
-\mathrm{Tr}(\hat{L}_s\rho_{\ms{c}}+\rho_{\ms{c}} \hat{L}^{\dagger}_{s}) (\hat{L}_r\rho_{\ms{c}}+\rho_{\ms{c}} \hat{L}^{\dagger}_{r})\\
-\mathrm{Tr}(\hat{L}_r\rho_{\ms{c}}+\rho_{\ms{c}} \hat{L}^{\dagger}_{r})(\hat{L}_s\rho_{\ms{c}}+\rho_{\ms{c}} \hat{L}^{\dagger}_{s})\\
+2\mathrm{Tr}(\hat{L}_r\rho_{\ms{c}}+\rho_{\ms{c}} \hat{L}^{\dagger}_{r})\mathrm{Tr}(\hat{L}_s\rho_{\ms{c}}+\rho_{\ms{c}} \hat{L}^{\dagger}_{s}) \rho_{\ms{c}}
\end{array}
\right)\left(\Delta W_r\Delta W_s-\delta_{r,s}\Delta t\right) , 
\end{eqnarray}
\end{widetext}
where we have written $\rho_{\ms{c}}$ and $\Delta \rho_{\ms{c}}$ as shorthand for $\rho_{\ms{c}}^{(n)}$ and $\Delta \rho_{\ms{c}}^{(n)}$. The Euler-Milstein increment neither guarantees the positivity nor the Hermiticity of the conditional density matrix. In practice, we have found that Rouchon's method \cite{Ami2011,Rou2015} provides more accurate state estimates with fewer time steps per period of time, and involves fewer calculations per time step than the Euler-Milstein method. For the cases modeled in this paper, typical savings in computational time are a factor of four or five for the increment (\ref{sme2}) over the increment (\ref{sme3}). 

As noted above, Rouchon's increment only guarantees first-order convergence and the positivity of the density matrix when all the operators $L_r$ for which $\eta_r > 0$ are mutually commuting. This condition of `commuting measurements' is fulfilled for our system, but it would be broken if $\eta_2$ and $\eta_3$ were nonzero. To examine the effect of breaking this condition we calculated approximate corrections to the stochastic integrals that would need to be added to Rouchon's increment to guarantee first-order convergence when $\eta_2 = \eta_3 = 1$~\cite{Klo1999}. These corrections were found to be at least an order of magnitude smaller than the errors shown in Fig.~\ref{fig:accuracy} below. This indicates that Rouchon's method may also perform favorably when the SME contains non-commuting measurements.  

The results presented in Sec.~\ref{sec:sec5} below were generated using Rouchon's method. The method was implemented using fifteen angular momentum states $J_z = -7,\ldots,+7$ and a minimum of 11 harmonic oscillator states for the optical tweezers (although slightly larger oscillator bases are required in some cases, see below). This gives a minimum of 165 states formed from the tensor product of 11 translational and 15 rotational states. The resulting conditional density matrix, $\rho_{\ms{c}}$, is a $165\times 165$ matrix with complex off-diagonal elements, and is more than 40 times larger than the $4\times 4$ two-qubit example studied in \cite{Rou2015}. 

We now compare the accuracy of Rouchon's method against the standard Euler-Milstein method, Eq.(\ref{sme3}), and display the results in Fig.~\ref{fig:accuracy}. To evaluate the accuracy of both methods we first generate a very accurate reference solution by using either integration method with a time-step of $\Delta t = T_{osc}/10^4$, in which $T_{osc}=2\pi/\omega_t$ is the period of the oscillator formed by the optical tweezers. We then perform simulations with both methods using a range of larger time-steps, and compare the resulting evolution of the density matrix, $\rho_{\ms{c}}(t)$, to that given by the reference solution. Denoting the latter by $\rho_0(t)$, we compare the solutions by calculating the `infidelity' between $\rho_{\ms{c}}$ and $\rho_0$. The infidelity is defined as $\varepsilon = 1-F$, where $F$ is the fidelity and is given by~\cite{Pet2004}
$$ 
F= F(\rho_{0},\rho_{\ms{c}}) = \left|{\rm Tr}\left[\sqrt{\sqrt{\rho_{\ms{c}}} \rho_{0} \sqrt{\rho_{\ms{c}}}}\right]\right|^2 . 
$$ 
Our measure of the error of a solution $\rho_{\ms{c}}(t)$ is the infidelity, $\varepsilon$, between $\rho_{\ms{c}}(t)$ and $\rho_0(t)$, averaged over the duration of the simulation. 

From Fig.~\ref{fig:accuracy} we see that Rouchon's method offers significantly higher accuracy than the Euler-Milstein method for larger time-steps. For our example, Rouchon's method produces an acceptably accurate solution using as few as 500 time steps per oscillator period, whereas the Euler-Milstein method requires at least this number merely to produce a solution that is reasonably stable, and requires approximately 2500-5000 time steps per oscillator cycle to provide a solution that does not significantly break the condition that $P=\mbox{Tr}[\rho^2] \leq 1$. As noted in \cite{Rou2015}, Rouchon's method automatically enforces this condition. 

We also found that for Rouchon's method the increment $\Delta\rho_{\ms{c}}$ was $8-10$ times faster to calculate  than that for the Euler-Milstein method. Nevertheless it is important to note that this speed does depend on the precise method used to implement the calculation, and the implementations used here may not have been optimal for either method.   
\begin{figure}[t] 
   \centering
   \includegraphics[width=1\hsize]{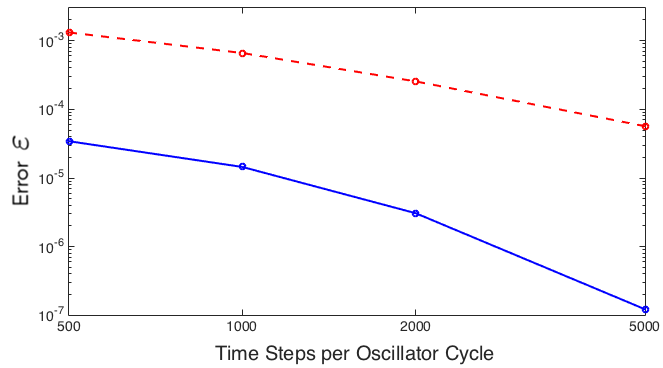} 
   \caption{Accuracy comparison for Rouchon's stochastic integration method (solid-blue) versus Euler-Milstein method (dash-red) for sphere radius $R=1.75$nm and $k = 0.005\omega_t$. Other parameter values for the optical tweezers and the dielectric sphere are given in the text.}
   \label{fig:accuracy}
\end{figure}

\section{\label{sec:sec5}NUMERICAL RESULTS}

The parameter space for our system is quite large, so we concentrate on the regime in which the interaction between the translational and rotational states is expected to be significant --- that is,  where the energy scales for the two degrees of freedom are comparable and any effect of the measurement interaction would be expected to be maximized. 

As discussed above, we expect the measurement to both entangle and correlate the position and orientation of the sphere in the conditioned state. By virtue of the continual extraction of information, the measurement will also purify the state and will drive both degrees of freedom with back-action noise -- noise that, in this case, is correlated between the degrees of freedom. This back-action noise will increase the energy associated with the sphere's motion, but we can expect that the addition of thermal noise from the unprobed environment will act to reduce the purity of the conditioned state and any correlations between the position and the orientation of the sphere. We now examine the entanglement, classical correlations, purity and energy of the conditioned state. 

As an indication of the level of entanglement between the translational and rotational degrees of freedom we use the `negativity' of the density matrix. The negativity is defined in the following way. Given a joint density matrix, $\rho_{\ms{AB}}$, for two degrees of freedom A and B, the partial transpose of $\rho_{\ms{AB}}$ with respect to A is denoted by $\rho_{\ms{AB}}^{\ms{T}(A)}$. The negativity of $\rho$ is then~\cite{Vid2002} 
\begin{align}
   {\cal N}(\rho)=\frac{||\rho^{\ms{T(A)}}||_{1} -1}{2} , 
\end{align}
in which $||\cdot ||_{1}$ is the trace norm.
 
The classical correlations between the two degrees of freedom can be quantified by the mutual information. Denoting the von Neumann entropy of a density matrix $\rho$ by $S(\rho)$, the mutual information may be written as  
\begin{align}
   I(\rho_{AB})=S(\rho_A)+S(\rho_B)-S(\rho_{AB}) , 
\end{align}
where $\rho_{\ms{AB}}$ is the joint density matrix and $\rho_{\ms{A}}$ (respectively $\rho_{\ms{B}}$) is the density matrix obtained from $\rho_{\ms{AB}}$ by taking the partial trace over B (respectively A).   

\begin{figure}[t] 
   \centering
   \includegraphics[width=1\hsize]{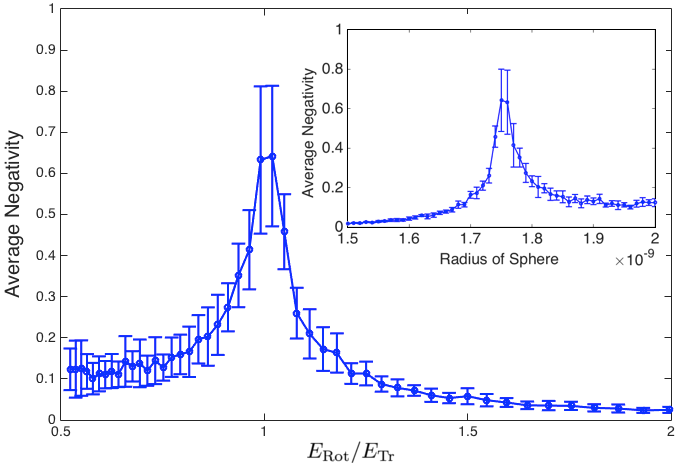} 
   \caption{Average Negativity for dielectric sphere versus ratio between the rotational and translational energies,  $E_{\ms{Rot}}/E_{\ms{Tr}}$ for $k = 0.005\omega_t$ and (inset) Average Negativity versus radius of the dielectric sphere $R$. Error bars shown are 1$\sigma$ fluctuations about the mean value. The mean and standard deviation were calculated in the steady state (after 100 oscillator cycles) and over 200 individual realizations for each point. Other parameter values for the optical tweezers and the dielectric sphere are given in the text.}
   \label{fig:radius}
\end{figure}

In the simulations, we start the system in the steady-state that the master equation possesses in the absence of the measurement. This means that the initial state is a tensor product state of the ground state for the rotational degree of freedom (that is, the zero eigenstate of $J_z$) and a thermal state for the translational degree of freedom. We set the temperature of the thermal state to be $T=5$ $\mu$K.

In Fig.~\ref{fig:radius} we plot the negativity generated between the two degrees of freedom by the measurement interaction in the absence of the thermal environment, where this negativity is averaged over all trajectories. While the entanglement may have an important role in distinguishing coherent quantum measurements from uncorrelated quantum measurements and classical measurements (see below), the amount of entanglement created by the measurement is relatively small in view of the number of states involved in the joint dynamics. Our interest in the negativity is not only as a measure of the entanglement, but as an indication of the size of the effect  of the coupling induced by the measurement on the dynamics of the sphere. Fig.~\ref{fig:radius} shows that the average negativity is maximized when the rotational and translational energy scales are comparable, as expected, and the error bars show that there is appreciable variability in the level of negativity, and hence entanglement, between the two degrees of freedom. As seen from the inset in Fig.~\ref{fig:radius}, the negativity is maximized when the radius of the dielectric sphere is around $R=1.75$ nm, and so we use this value in what follows. These were obtained using Rouchon's method with 500 integration steps per oscillator period $T_{osc}$.  

\begin{figure}[t] 
\leavevmode\includegraphics[width=1\hsize]{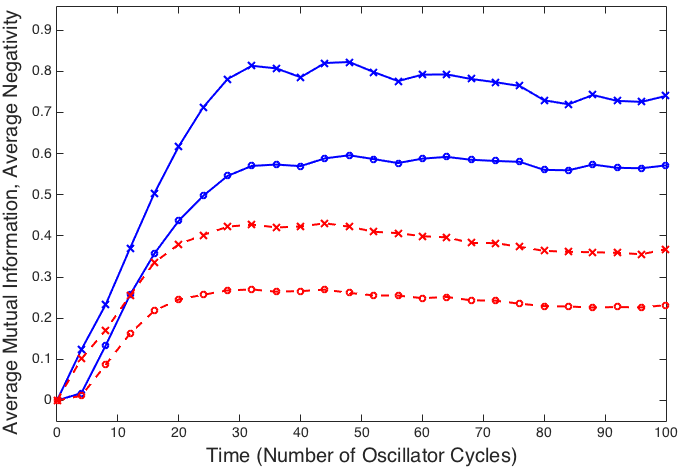} 
\caption{Average mutual information (crosses) and average negativity (circles) for a dielectric sphere as a function of time for $k = 0.005\omega_t$. Sphere radius is fixed to be $R=1.75$nm so that $E_{\ms{Rot}}/E_{\ms{Tr}}\simeq 1$, and averages are calculated over 500 realizations with (dash-red) and without the thermal environment (solid-blue).} 
\label{fig:negativity} 
\end{figure}

In Fig.~\ref{fig:negativity} we plot both the average negativity and the average mutual information between the two degrees of freedom as a function of time. We see that the quantum correlations, indicated by the negativity, and the classical correlations are essentially proportional to each other. 

\begin{figure}[t] 
\leavevmode\includegraphics[width=1\hsize]{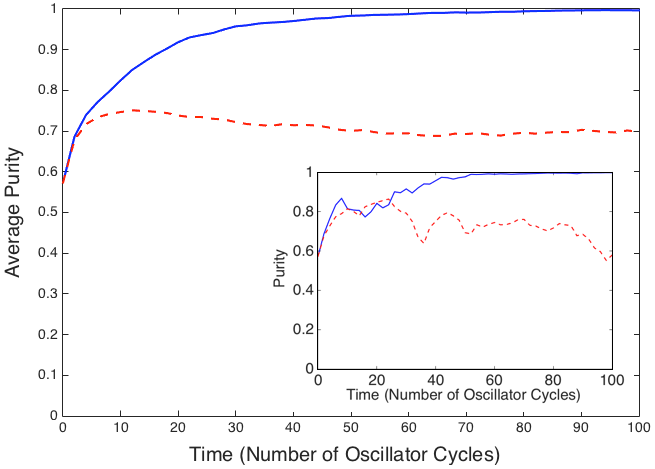} 
\caption{Average Purity for dielectric sphere as a function of time for $k = 0.005\omega_t$. Sphere radius is fixed to be $R=1.75$nm so that $E_{Rot}:E_{Tr}\simeq 1$. Averages are calculated with (dash-red) and without the thermal environment (solid-blue). Insert shows the purity of an example trajectory, with and without the thermal environment} 
\label{fig:purity} 
\end{figure}

We consider next the purity of the evolving conditional density matrix. Since the translational degree of freedom starts in a thermal (mixed) state, in the absence of the thermal noise the action of the measurement is to purify this state. Without the thermal noise, for all realizations of the measurement process the system tends towards a purity of unity, but the evolution is stochastic. We show this behavior in Fig.~\ref{fig:purity}. Adding the coupling to the dissipative thermal environment via $\hat{L}_2$ and $\hat{L}_3$ reduces the asymptotic value of the average purity to $\bar{P} = 0.7$, indicating that information regarding the evolution of the system is lost to the environmental degrees of freedom. The purity of the individual realizations (or quantum trajectories) is still stochastic, but not limited to values around $P =0.7$. The purity of the individual realizations can approach $P =1$, even if the value does not remain there for very long (see Fig.~\ref{fig:purity} insert). 

\begin{figure}[t] 
   \centering
   \includegraphics[width=1\hsize]{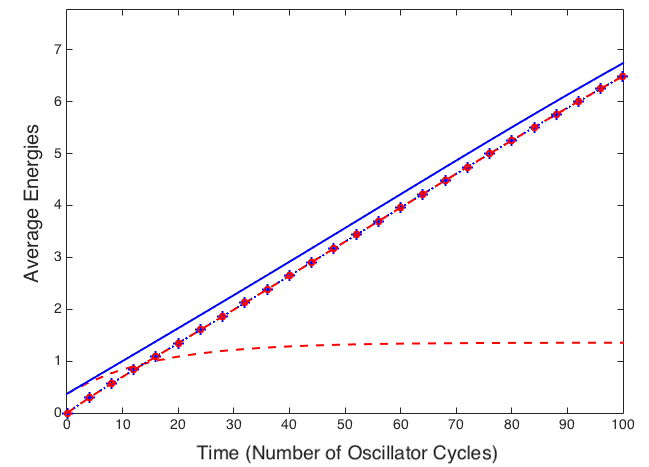} 
   \caption{Average Energy for dielectric sphere as a function of time for $k = 0.005\omega_t$. Sphere radius is fixed to be $R=1.75$nm so that $E_{Rot}:E_{Tr}\simeq 1$. Average energies are calculated with the thermal environment -- translational (dash-red) and rotational (dash-red-circles) and without the thermal environment -- translational (solid-blue) and rotational (dotted-blue-plusses).}
   \label{fig:energy}
\end{figure}

In Fig.~\ref{fig:energy} we plot the average energy as a function of time for the cases shown in Fig.~\ref{fig:purity}. The average energy, $\bar{E}$ is especially simple to calculate because it is a linear function of the density matrix. Instead of calculating trajectories using the SME we need merely integrate the master equation to obtain the averaged state, $\rho$, and use $\bar{E}=\mbox{Tr}[\rho \hat{H}]$. From Fig.~\ref{fig:energy} we see that the average energy of each subsystem increases due to the effect of the measurement operator $\hat{x} =( \hat{X}+R\hat{S})$. The partial localization of the center of mass position through the use of the $\hat{X}$ operator, and an associated reduction in the variance of the state, causes the energy to increase as the width of state is reduced relative to that of the initial thermal mixed state. A similar process occurs for the rotational states due to the partial localization of $\sin\theta$. In the absence of the thermal environment, the energy for both the translational and the rotational degree of freedom grow linearly in time. The addition of the thermal environment limits the energy growth in the translational degree of freedom, but not the rotational degree of freedom. This produces a steady state on average for the oscillator; where the energy increase due to the localizing effect of the measurement is balanced by the dissipative effect of the environment. The measurement strength and parameters associated with the thermal environment (the strength of the coupling parameter $k$, and the temperature $T$ and quality factor $Q$) are such that the system would be expected to approach a steady state after between 50 and 100 oscillator cycles -- the average purity, the average negativity and the translational mode energy all reach a steady state value after around 50 cycles of the simulation.  

The dissipative effect of the environmental operators also has a numerical benefit since it limits the number of states required in the simulation. As the energy increases, more and more oscillator states are required to solve the SME. In fact, to produce the linear energy growth for the translational degree of freedom shown in Fig.~\ref{fig:energy} requires up to 25 oscillator states in the numerical basis (giving a $375\times 375$ density matrix). The use of thermal environment operators allows the truncation of the oscillator basis to 11 states without losing the accuracy of the numerical solutions. The energy growth of the rotational states used in the simulations would also ultimately be limited by the finite basis used in the calculations. In practice, it could itself be coupled to additional environmental operators. However, the simulation of a thermal environment for such a subsystem is more problematic because the energy level spacings are not uniform and it would require coupling the states to an environment which could contain a number of different frequencies and dissipation mechanisms. 

\section{\label{sec:sec6}DISCUSSION}

The solution of the SME provides, in the form of a density matrix, a complete description of the observer's knowledge of the sphere's motion that she/he obtains from the continuous measurement. In the presence of a thermal environment, the density matrix contains some classical uncertainty (it will in general be mixed), but it represents everything that the observer knows about the motion of the sphere at each point in time for a given measurement record, including any classical and quantum correlations. The reconstruction of quantum trajectories from measurement records has been demonstrated  experimentally, albeit for systems with a smaller number of basis states~\cite{Mur2013,Web2014}. The reconstruction of a quantum trajectory from an experimental measurement record using Rouchon's method has also been demonstrated and is found to be in good agreement with quantum tomography performed on the final state~\cite{Six2015}. Given these developments, it is likely that more complex quantum states of individual systems will be reconstructed by solving the SME in the near future. The density matrices obtained in this way will provide estimates of the purity and the quantum and classical correlations between different degrees of freedom. A benefit of this approach would be the ability to use the state estimates as part of a feedback loop to control the behavior of the quantum system~\cite{Bel1980,Wis2010}. 

\begin{figure}[t] 
\leavevmode\includegraphics[width=1\hsize]{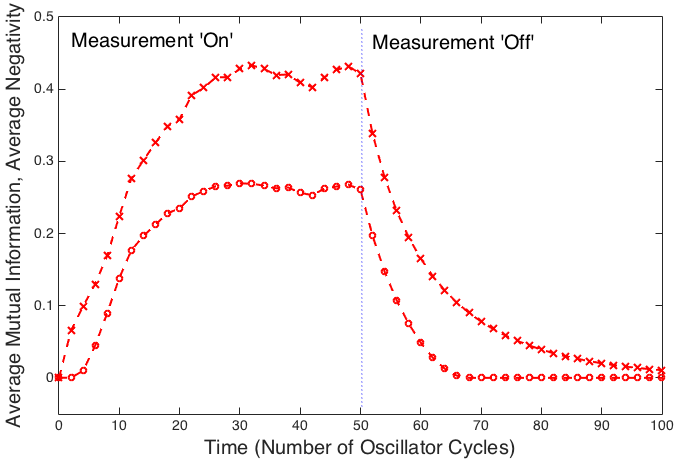} 
\caption{Average mutual information (crosses) and average negativity (circles) for a dielectric sphere as a function of time with $k = 0.005\omega_t$ for $t<50T_{osc}$ (measurement `on') and $k = 0$ for $t > 50T_{osc}$ (measurement `off'). Sphere radius is fixed to be $R=1.75$nm so that $E_{Rot}:E_{Tr}\simeq 1$, and averages are calculated over 500 realizations with the thermal environment.} 
\label{fig:negativity2} 
\end{figure}

In the scenario we have considered here, we did not apply classical forces to the sphere specifically to control its motion. As previously noted, it is possible to introduce an optical cavity which can be used to manipulate the behavior of the dielectric sphere -- for example to implement feedback cooling~\cite{Rom2011b,Yin2013}. In our description of the trapped sphere we have removed the explicit optical cavity to simplify the system and to make the integration of the SME tractable. The inclusion of a cavity could also provide a direct measurement of position, and this could affect the combined measurement of the position and rotation by localizing $X$ independently. In the absence of a cavity field, one can apply a feedback process by adjusting the measurement strength, $k$, based on the current motion of the sphere. For example, if the measurement interaction is turned off at some point during the evolution, the two degrees of freedom will evolve independently -- ignoring the very weak indirect coupling noted above~\cite{Rom2011b}. Since the translational mode is coupled to a thermal bath, removing the measurement will cause the translational state to relax back to a thermal mixed state. In such situations, any quantum correlations between the translational and rotational states will decay very rapidly and classical correlations will decay more slowly. This can be seen Fig.~\ref{fig:negativity2}, where the average negativity vanishes within 15-20 oscillator cycles and the mutual entropy decays slowly over more than 50 cycles. Eventually, the two degrees of freedom will evolve towards a separable state, in which the translational state is a thermal mixed state and the rotational state is `frozen' at the point that the measurement was turned off. Typically, the rotational states generated by the measurement interaction are equal superpositions of $\pm j_z$ eigenstates of the $\hat{J}_z$ operator; the average value of $j_z$ grows with time resulting in the linear energy growth seen in Fig.~\ref{fig:energy}. While the final state could be controlled using a simple feedback process in which the measurement is turned on and off, this does not seem like a very efficient procedure since one would have to wait for the final state to occur stochastically (preferably with high purity) before turning off the measurement. In view of this, we note that there may be more sophisticated ways to modulate the measurement strength that would produce desired states more predictably or efficiently~\cite{Jacobs10}. 

In addition to feedback control, the measurement-induced coupling we have considered could have applications in the verification of quantum mechanics in extended systems (i.e. systems that are `large' compared to single atoms and ions). In particular, the use of a continuous measurement model and the derivation of an accurate quantum trajectory require that the measurement operator accurately reflects the effect of the measurement on the quantum system. If the measurement model is inaccurate, or the Hamiltonian evolution is incorrect, the estimated state $\rho_{\ms{c}}$ derived from the classical measurement results $y(t)$ will deviate from the true state of the system. One can test the deviation by using tomography to reconstruct the final state. Experiments to reconstruct quantum trajectories have already demonstrated such agreement for `small' systems~\cite{Mur2013,Web2014,Six2015}. 

The measurement of the operator $\hat{x} = \hat{X} + R\hat{S}$ is a quantum mechanical process, relying on the coherence of the coupled measurement. The energy growth noted in Fig.~\ref{fig:energy} is a result of the quantum back-action associated with the measurement of the two degrees of freedom, but it does not show whether the measurement was coherent or not. There is more than one way to make a simultaneous measurement of two observables, or equivalently, more than one way to model a simultaneous measurement. The alternative models include: 1) making separate measurements of each observable  -- $L_X \propto \hat{X}$ and $L_{RS}  \propto R\hat{S}$ -- thus obtaining two separate measurement records -- $y_X(t)$ and $y_{RS}(t)$ respectively -- and independent state updates, and 2) making the same separate measurements but combining the two measurement records so that the observer only has access to the single record $\tilde{y}(t) = y_X(t)+y_{RS}(t)$. The evolution of $\rho_{\ms{c}}$ predicted by the SME would be different for each of the three different measurement models: (i) $\hat{x} = \hat{X} + R\hat{S} \rightarrow y(t)$, (ii) $\hat{X}, R\hat{S}\rightarrow y_X(t), y_{RS}(t)$, and (iii) $\hat{X}, R\hat{S}\rightarrow \tilde{y}(t) = y_X(t)+y_{RS}(t)$. It could be rigorously verified that an experiment was performing a coherent measurement of $\hat{x}$ (model (i)), for example, by comparing the final states predicted by the three models and comparing these, via hypothesis testing~\cite{Tsang12, Tsang13}, to the state obtained experimentally using tomography. All three cases would show the same energy growth but the quantum and classical correlations would be different. Case (i) would show quantum and classical correlations -- as discussed above. For case (ii), the respective quantum states would purify under the action of the measurements, but the subsystem states would remain separable and the correlations (including the negativity) would be zero. By contrast, case (iii) would predict classical correlations between the translational and rotational subsystems, but not quantum correlations, i.e. the negativity of the reconstructed state would be zero. A similar situation would occur with purely classical measurements. Classical correlations could arise from a combined measurement on a single system, but quantum correlations would not occur. The appearance of entanglement would be evidence of quantum mechanical behavior and a coherent measurement process. A non-zero value for the negativity from quantum tomography of the final state would provide such evidence, even if the value of the negativity was relatively small. These differences between quantum measurements, and the ability to differentiate between different types of correlated and uncorrelated measurements, may provide interesting and non-trivial litmus tests for alternative macroscopic quantum theories \cite{Bas2003,Bas2013,Ber2015}. 

Lastly, we consider the approximations we have used to simplify the dynamics of the dielectric sphere to render the stochastic integration of the SME tractable: the reduction to one translational and one rotational degree of freedom. The key to each of these approximations is a separation of energy scales for the various excitations. Treating the optical trap/tweezers as a one-dimensional harmonic potential is fairly standard \cite{Rom2011b,Che2012,Yin2013}, as is neglecting the very weak coupling between the rotational and translational states \cite{Rom2011b}. More problematic is the reduction of the rotational degrees of freedom to one-dimension. The approximation we have used is based on a mathematical limit \cite{Car1968}. While it allows a simple angular momentum basis to be used in calculations, this basis may not completely represent the three-dimensional nature of rotations of a dielectric sphere. Nevertheless, spheres are not the only type of extended structures to be trapped in optomechanical levitation experiments. Other asymmetric structures, such as rods or tops, would provide a physical mechanism that separates the energy scales of the different angular momentum states \cite{Sti2016,Kuh2015,Hoa2016} and could allow the same type of measurements to be performed. The internal states, however, may be even more problematic than the rotational states. The vibrational states of the sphere have a much higher energy than those of the center-of-mass motion \cite{Rom2011b} but there are other issues that already present significant challenges in experimental systems. A trapped dielectric sphere tends to scatter photons which causes heating of the motional states and the material of the crystal itself \cite{Yin2013}. This scattering could itself act as a form of position measurement and thus affect the dynamics induced by the measurement proposed in this paper. However, given the present interest in demonstrating the quantum behavior of macroscopic objects, and in probing the quantum/classical boundary, addressing these issues is a focus of current experimental efforts. 

\section{\label{sec:sec7}CONCLUSIONS}

In this paper we have studied a simple model of a novel form of measurement for an optomechanically levitated dielectric sphere. This measurement exploits the spatially extended nature of the sphere to provide a coupling between the rotational and translational states of the system through the localization of a spot on the surface of the sphere. 

The stochastic master equation that describes the evolution of the system under the continuous measurement was solved numerically using a method proposed by Rouchon and collaborators that was specially designed for these types of equations~\cite{Ami2011,Rou2015}. In doing so we have confirmed, for systems much larger than those previously considered, that Rouchon's method provides significant computational advantages compared to the standard Euler-Milstein method. 

We have shown that the effect of the continuous measurement on the sphere was to generate correlations between the two degrees of freedom. We found that the growth of energy in the translational degree of freedom was limited by the introduction of a thermal environment, which reduced but did not remove the quantum and classical correlations between the two subsystems. We also discussed briefly the possibility of using the measurement in a feedback loop to control the motion. 

The effect of the measurement proposed in this paper is quantum mechanical in nature, arising from the back-action of the measurement on the evolution of the system (energy growth) and the correlated/coherent nature of the measurement interaction. As such, the interaction provides a novel method to explore the quantum-classical interface and possible deviations from standard quantum mechanics in large systems. 

\begin{center}
	\textit{Acknowledgments}
\end{center}
\noindent JFR would like to thank Jared Cole, James Babington, Peter Barker, Mark Everitt and Pierre Rouchon for useful discussions during the preparation of this paper. JFR would also like to thank the organizers of the UCL/EPSRC workshop on `Quantum control of levitated optomechanics' (Pontremoli, May 2016).

\bibliographystyle{apsrev}

\end{document}